\documentclass[11pt]{article}

\usepackage[reqno]{amsmath}
\usepackage{epsfig}
\usepackage{array}
\usepackage{float}

\begin{document}


\def\ltap{\ \raisebox{-.4ex}{\rlap{$\sim$}} \raisebox{.4ex}{$<$}\ }
\def\gtap{\ \raisebox{-.4ex}{\rlap{$\sim$}} \raisebox{.4ex}{$>$}\ }
\newcommand{\deltaatm}{\mbox{$\Delta  m^2_{\mathrm{atm}} \ $}}
\newcommand{\deltasol}{\mbox{$ \Delta  m^2_{\odot} \ $}}
\newcommand{\utre}{\mbox{$|U_{\mathrm{e} 3}|$}}
\newcommand{\betabeta}{\mbox{$(\beta \beta)_{0 \nu}  $}}
\newcommand{\mefff}{\mbox{$ < \! m  \! > $}}
\newcommand{\meff}{\mbox{$\left|  < \! m  \! > \right| \ $}}
\newcommand{\hbeta}{$\mbox{}^3 {\rm H}$ $\beta$-decay \ }
\newcommand{\eV}{\mbox{$ \  \mathrm{eV} \ $}}
\newcommand{\deltatre}{\mbox{$ \ \Delta m^2_{32} \ $}}
\newcommand{\deltadue}{\mbox{$ \ \Delta m^2_{21} \ $}}
\newcommand{\ueuno}{\mbox{$ \ |U_{\mathrm{e} 1}|^2 \ $}}
\newcommand{\uedue}{\mbox{$ \ |U_{\mathrm{e} 2}|^2 \ $}}
\newcommand{\uetre}{\mbox{$ \ |U_{\mathrm{e} 3}|^2  \ $}}


\hyphenation{par-ti-cu-lar}
\hyphenation{ex-pe-ri-men-tal}
\hyphenation{dif-fe-rent}
\hyphenation{bet-we-en}
\hyphenation{mo-du-lus}
\hyphenation{}

\rightline{Ref. SISSA 81/2001/EP}
\rightline{October 2001}
\vskip 0.6cm
\begin{center}
{\bf
Searching for the CP-Violation Associated with Majorana Neutrinos\\ 
} 

\vspace{0.3cm} 
S. Pascoli$^{(a,b)}$,~~S. T. Petcov$^{(a,b)}$~
\footnote{Also at: Institute of Nuclear Research and
Nuclear Energy, Bulgarian Academy of Sciences, 1784 Sofia, Bulgaria}
and~~L. Wolfenstein$^{(c)}$ 

\vspace{0.2cm}   
{\em $^{(a)}$ Scuola Internazionale Superiore di Studi Avanzati, 
I-34014 Trieste, Italy\\
}
\vspace{0.2cm}   
{\em $^{(b)}$ Istituto Nazionale di Fisica Nucleare, 
Sezione di Trieste, I-34014 Trieste, Italy\\
}
\vspace{0.2cm}   
{\em $^{(c)}$ Department of Physics, Carnegie-Mellon 
University, Pittsburgh, PA 34014, U. S. A.\\
}
\end{center}

\vskip 0.4cm
\begin{abstract}
The effective Majorana mass 
which determines the rate of the
neutrinoless double beta
(\betabeta-)decay, \meff, is considered 
in the case of three-neutrino mixing 
and massive Majorana neutrinos. 
Assuming a rather precise determination of the 
parameters characterizing the 
neutrino oscillation solutions
of the solar and atmospheric neutrino
problems has been made,
we discuss 
the information a measurement of 
$\meff \gtap (0.005 - 0.010)$ eV 
can provide on the value of the 
lightest neutrino mass
and on the CP-violation 
in the lepton sector.
The implications of 
combining a measurement of \meff
with future measurement
of the neutrino mass $m_{\nu_e}$ in 
$^3$H $\beta-$decay experiments
for the possible determination  
of leptonic CP-violation
are emphasized.

\end{abstract}
\vfill
\newpage
\vspace{-0.3cm}
\section{Introduction} 
\vspace{-0.2cm}
\indent  Experiments on atmospheric  
and solar neutrinos have 
produced convincing evidence of neutrino oscillations 
\cite{SKatm98,Cl98,Fu96,SAGE,GALLEXGNO,SKsol,SNO1}. 
Ongoing and planned experiments,
including long baseline ones, aim to determine 
the parameters for these oscillations. 
Assuming mixing of only three neutrinos, these are the 
magnitudes of the elements of the 
$3\times 3$ unitary lepton mixing matrix - the
Maki-Nakagawa-Sakata-Pontecorvo 
(MNSP) mixing matrix \cite{BPont57,MNS62}, 
a CP-violating phase, and the mass-squared 
difference parameters, say, $\Delta m^2_{31}$
and  $\Delta m^2_{21}$.
In principle, long baseline experiments 
at neutrino factories
can distinguish the alternatives of 
a (i) hierarchical neutrino mass spectrum and
of (ii) neutrino mass spectrum 
with inverted hierarchy \cite{LBLhih1,LBLhih2}.
If we number (without loss of generality) 
the neutrinos with definite mass
in such a way that $m_1 < m_2 < m_3$,
case (i) corresponds to 
$\Delta m^2_{31} \equiv \Delta m^2_{atm} \gg 
\Delta m^2_{21} \equiv \Delta m^2_{sol}$, while 
in case (ii) we have 
$\Delta m^2_{31} \equiv \Delta m^2_{atm} \gg 
\Delta m^2_{32} \equiv \Delta m^2_{sol}$,
where $\Delta m^2_{atm}$ and $\Delta m^2_{sol}$ 
are the values of the neutrino mass-squared 
differences inferred from the atmospheric 
and solar neutrino data.
These experiments cannot determine, 
however, the actual neutrino masses, that is, 
the value of the lightest 
neutrino mass $m_1$. Furthermore,  
assuming the massive neutrinos 
are Majorana particles, as we will in 
this paper, there are two more parameters, two
Majorana CP-violating phases, associated with the
MNSP mixing matrix \cite{BHP80} (see also \cite{Kobz80}). 
The neutrino oscillation 
experiments cannot provide information
on the Majorana CP-violating phases
\cite{BHP80,LPST87} as well. 
This paper is concerned 
with the prospects and
problems in determining or constraining 
these three parameters, assuming
the others have been well determined. 
The mass $m_1$ is of interest, e.g.,
in cosmology since massive neutrinos at 
present are the only non-baryonic 
dark matter constituents known. 
Knowing the neutrino mass spectrum is 
fundamental for understanding the 
origin of the neutrino masses and mixing.
The Majorana CP-violating 
phases indicate the relation between 
CP violation and lepton number violation; a major
goal is to identify any possibility 
of detecting this CP-violation.

\vspace{-0.3cm}
\section{Neutrinoless Double $\beta-$Decay and 
$^3$H $\beta-$Decay Experiments} 
\vspace{-0.2cm}
\indent The process most sensitive to the 
existence of massive Majorana neutrinos
(coupled to the electron 
in the weak charged lepton current)
is the neutrinoless double beta 
($\betabeta-$) decay (see, e.g., \cite{bb0nu1,bb0nu2}). 
If the $\betabeta-$ decay is generated 
{\it only} by the 
left-handed (LH) charged current weak interaction
through the exchange of virtual 
massive Majorana neutrinos, 
the probability amplitude of this process 
is proportional 
in the case of Majorana 
neutrinos having masses not 
exceeding a few MeV 
to the so-called 
``effective Majorana mass parameter'',
\meff (see, e.g., \cite{BiPet87}).
A large number of experiments are searching  for 
$\betabeta-$decay of different 
nuclei at present (a rather complete list 
is given in \cite{bb0nu2}). 
No indications that \betabeta-decay 
takes place have been found so far.
A stringent constraint on the 
value of the effective Majorana mass \meff
was obtained in the $^{76}$Ge Heidelberg-Moscow 
experiment \cite{76Ge00}: 
\begin{equation}
\meff < 0.35 \ \mathrm{eV},~~~90\%~{\rm C.L.}
\label{76Ge0}
\end{equation}
\noindent Taking into account a factor of 3 
uncertainty associated with the calculation of the relevant
nuclear matrix element (see, e.g., \cite{bb0nu1,bb0nu2})
we get
\begin{equation}
\meff < (0.35 \div 1.05) \ \mathrm{eV},~~~90\%~{\rm C.L.}
\label{76Ge00}
\end{equation}
\noindent The IGEX collaboration  
has obtained \cite{IGEX00}:
\begin{equation}
\meff < (0.33 \div 1.35) \ \mathrm{eV},~~~90\%~{\rm C.L.}
\label{IGEX00}
\end{equation}

  A sensitivity to $\meff \sim 0.10$ eV
is foreseen to be reached in the currently operating 
NEMO3 experiment \cite{NEMO3},
while the next generation 
of $\betabeta-$decay experiments CUORE,
EXO, GENIUS, MOON \cite{CUORE,GENIUS,EXO,MOON},
aim at reaching a sensitivity 
to values of $\meff \sim 0.01$ eV,
which are considerably smaller than the 
presently existing most stringent upper
bounds (\ref{76Ge00}) and (\ref{IGEX00}).

 The results of the 
\hbeta experiments
studying the electron spectrum, which
measure the electron (anti-)neutrino 
mass  $m_{\nu_e}$,
are of fundamental importance, in particular,
for getting information
about the neutrino mass spectrum.
The Troitzk~\cite{MoscowH3} and 
Mainz~\cite{Mainz} experiments
have provided stringent 
upper bounds on $m_{\nu_e}$: 
\begin{eqnarray}
m_{\nu_e}  <  2.5 \eV~~\cite{MoscowH3},\ \ 
m_{\nu_e} <  2.9  \eV~~~\cite{Mainz}~~~~~~~~~~~~~~~(95\%~{\rm C.L.}).
\label{H3beta}
\end{eqnarray}

\noindent There are prospects to 
increase substantially 
the sensitivity  
of the \hbeta experiments
and probe the region of
values of $m_{\nu_e}$
down to $m_{\nu_e} \sim (0.3 - 0.4)$ eV
\cite{KATRIN}~  
\footnote{Cosmological and astrophysical data provide
information on the sum of the neutrino masses.
The current upper bound reads (see, e.g., \cite{Cosmo}
and the references quoted therein):
$ \sum_{j} m_{j}  \ltap  5.5$ eV.
\noindent The future experiments MAP and PLANCK 
may be sensitive to \cite{MAP}
$\sum_{j} m_{j}  \cong  0.4$ eV.
}
(the KATRIN project).

  It is difficult to overestimate the importance of the 
indicated future $\betabeta-$decay and $^3$H 
$\beta-$decay experiments for the studies 
of the neutrino mixing: these are the only 
feasible experiments which can provide 
information on the neutrino mass spectrum and on 
the nature of massive neutrinos.
Such information
cannot be obtained \cite{BHP80,LPST87}, 
as we have indicated, 
in the experiments
studying neutrino oscillations.
The measurement of
$\meff \gtap 0.02$ eV and/or of 
$m_{\nu_e} \gtap 0.4$ eV 
can give information, in particular, on 
the type of neutrino mass spectrum
\cite{BGGKP99,BPP1,BPP2}. 
As we will discuss, it is only by combining a value of 
\meff and a value of, or a sufficiently 
stringent upper limit on, $m_{\nu_e}$ one might hope to detect
Majorana CP-violation.

\vspace{-0.3cm}
\section{A Brief Summary of the Formalism} 
\vspace{-0.2cm}
\indent As it is well known, 
the explanation of the atmospheric and 
solar neutrino data in terms of 
neutrino oscillations requires
the existence of 3-neutrino mixing
in the weak charged lepton 
current: 
\begin{equation}
\nu_{l \mathrm{L}}  = \sum_{j=1}^{3} U_{l j} \nu_{j \mathrm{L}},
\label{3numix}
\end{equation}
\noindent where $\nu_{lL}$, $l  = e,\mu,\tau$,
are the three left-handed flavour 
neutrino fields,
$\nu_{j \mathrm{L}}$ is the 
left-handed field of the 
neutrino $\nu_j$ having a mass $m_j$
and $U$ is 
the MNSP neutrino mixing matrix \cite{BPont57,MNS62}. 
If $\nu_j$ are Majorana neutrinos
with masses not exceeding few MeV,
as will be assumed in what follows, 
the effective Majorana mass \meff
of interest can be expressed 
in the form
\begin{equation}
\meff = \left| m_1 |U_{\mathrm{e} 1}|^2 
+ m_2 |U_{\mathrm{e} 2}|^2~e^{i\alpha_{21}}
 + m_3 |U_{\mathrm{e} 3}|^2~e^{i\alpha_{31}} \right|
\label{effmass2}
\end{equation}
\noindent where 
$\alpha_{21}$ and $\alpha_{31}$ 
are the two Majorana CP-violating phases
\footnote{We assume that 
the fields of the 
Majorana neutrinos $\nu_j$ 
satisfy the Majorana condition:
$C(\bar{\nu}_{j})^{T} = \nu_{j},~j=1,2,3$,
where $C$ is the charge conjugation matrix.}
\cite{BHP80} (see also \cite{Kobz80}).
If CP-invariance holds, 
one has \cite{LW81,BNP84,Kayser84}
$\alpha_{21} = k\pi$, $\alpha_{31} = 
k'\pi$, $k,k'=0,1,2,...$. In this case 
\begin{equation}
\eta_{21} \equiv e^{i\alpha_{21}} = \pm 1,~~~
\eta_{31} \equiv e^{i\alpha_{31}} = \pm 1 ,
\label{eta2131}
\end{equation}
\noindent represent the relative 
CP-parities of the neutrinos 
$\nu_1$ and $\nu_2$, and 
$\nu_1$ and $\nu_3$, respectively. 

  The quantities relevant for eq. (\ref{effmass2})
to be determined 
in neutrino oscillation
experiments in the case of 
three-neutrino mixing
are \deltaatm , \deltasol, 
the mixing angle, $\theta_{\odot}$, 
constrained by the solar neutrino data,
and the mixing angle, $\theta$,
determined from the probability
that the atmospheric neutrino oscillations
involve $\nu_e$. At present $\theta$
is limited by the
data from the CHOOZ \cite{CHOOZ} and 
Palo Verde \cite{PaloV} experiments,
but in the future it should be determined,
e.g., in long baseline neutrino 
oscillation experiments \cite{MINOS,LBLhih2,LBL3}.

  We can number (without loss 
of generality) the neutrino masses 
in such a way that $m_1 < m_2 < m_3$.
The neutrino masses $m_{2,3}$ can be expressed 
in terms of the lightest neutrino mass $m_1$
and, e.g., $\sqrt{\Delta m^2_{21}}$ 
and $\sqrt{\Delta m^2_{32}}$ (see, e.g., \cite{FV99,Poles00,KPS00}):
\begin{equation} 
m_{2} = \sqrt{m^2_{1} +  \Delta m^2_{21}}~,
\label{m2}
\end{equation}
\begin{equation} 
m_{3} = \sqrt{m^2_{1} + \Delta m^2_{21} +  \Delta m^2_{32}}~.
\label{m3}
\end{equation}

%
\noindent  For \deltaatm
inferred from the neutrino 
oscillation interpretation of the 
atmospheric neutrino data
we have:
\begin{equation} 
\deltaatm = \Delta m^2_{31} = \Delta m^2_{21} +  \Delta m^2_{32}~,
\label{dmatm0}
\end{equation}
%
\noindent In the case of normal neutrino mass hierarchy, 
\begin{equation}
\deltasol \equiv \deltadue,
\label{dmsolnh}
\end{equation}
%
\noindent and 
\begin{equation}
|U_{\mathrm{e} 1}| = \cos \theta_{\odot} \sqrt{1 - |U_{\mathrm{e} 3}|^2},
~~|U_{\mathrm{e} 2}| = \sin \theta_{\odot} \sqrt{1 - |U_{\mathrm{e} 3}|^2},
~~~|U_{\mathrm{e} 3}|^2 = \sin^2\theta.
\label{Ue123}
\end{equation}
%
\noindent For the inverted neutrino 
mass hierarchy one has \cite{BGKP96}: 
\begin{equation}
\deltasol \equiv \deltatre,
\label{dmsolih}
\end{equation}
%
\noindent and
\begin{equation}
|U_{\mathrm{e} 2}| = \cos \theta_{\odot} \sqrt{1 - |U_{\mathrm{e} 1}|^2},
~~|U_{\mathrm{e} 3}| = \sin \theta_{\odot} \sqrt{1 - |U_{\mathrm{e} 1}|^2},
~~~|U_{\mathrm{e} 1}|^2 = \sin^2\theta.
\label{Ue231}
\end{equation}
%

   In our analysis we will consider values 
of $m_1$ varying from 0 to 2.9 eV - the 
upper limit from the \hbeta data, eq.  (\ref{H3beta}).
As $m_1$ increases from 0, the three neutrino masses 
get closer in magnitude  
\footnote{For the values of \deltaatm
obtained in \cite{SKatm00}, 
one has neutrino mass spectrum with hierarchy
(with partial hierarchy) 
or with inverted hierarchy
(partial inverted hierarchy) for \cite{BPP1} 
$m_1 \ll 0.02$ eV 
($0.02~{\rm eV}\ltap m_1 \leq 0.2$ eV).}. 
For $m_1 > 0.2$ eV, the neutrino masses 
are quasi-degenerate and the
differences between the cases of hierarchical spectrum
and the spectrum with inverted hierarchy 
essentially disappear. 

  Given the values of $\deltasol$, $\theta_{\odot}$, 
$\deltaatm$ and of $\theta$, the effective Majorana mass
\meff depends, in general, on three parameters:
the lightest neutrino mass $m_1$ and on the two 
CP-violating phases $\alpha_{21}$ and $\alpha_{31}$.
It depends also on the ``discrete ambiguity''
expressed in eqs. (\ref{dmsolnh}) -
(\ref{Ue231}) and related 
to the two possible types of 
neutrino mass spectrum - 
the hierarchical and that   
with inverted hierarchy.
As is obvious from eqs. (\ref{m2}) - 
(\ref{dmsolnh}) and (\ref{dmsolih}), 
the knowledge of $m_1$ would allow to 
determine the neutrino mass spectrum.

   In the discussion which follows we use 
the best fit value for 
\deltaatm, obtained in the analysis 
of the atmospheric neutrino data in 
\cite{SKatm00}, 
\begin{equation} 
(\deltaatm)_{BFV} = 2.5\times 10^{-3}~{\rm eV^2}~.
\label{bfdmatm}
\end{equation}
%
\noindent In what regards the parameters
\deltasol and $\theta_{\odot}$, in most of 
the discussion we assume they lie in the region
of the large mixing angle
(LMA) MSW solution of the solar neutrino problem,
although we comment briefly on how our 
conclusions would change in the cases of the 
LOW - quasi-vacuum oscillation
(LOW-QVO) solution and of the
small mixing angle (SMA) MSW solution.
The most recent analyses 
\cite{FogliSNO,ConchaSNO,GoswaSNO,GiuntiSNO} 
show that the current solar neutrino data,
including the SNO results,
favor the LMA MSW and the LOW-QVO solutions.
To illustrate our discussion and conclusions
we use the best fit value of 
\deltasol found in \cite{ConchaSNO},
\begin{equation} 
(\deltasol)_{BFV} = 4.5\times 10^{-5}~{\rm eV^2}~,
\label{bfdmsol}
\end{equation}
%
\noindent three values of 
$\cos2\theta_{\odot}$ from the 
LMA solution region~\footnote{
In our further discussion we assume
$\cos2\theta_{\odot} \geq 0$,
which is favored by the analyses 
of the solar neutrino data 
\cite{FogliSNO,ConchaSNO,GoswaSNO,GiuntiSNO}.
The modification of the relevant formulae
and of the results 
in the case $\cos2\theta_{\odot} < 0$ 
is rather straightforward.}, and two values 
of the mixing angle $\theta$,
constrained by the CHOOZ 
and Palo Verde data. 

\vspace{-0.3cm}
\section{Constraining or Determining the Lightest Neutrino Mass 
$m_1$ and/or the Majorana CP-Violating Phases}
\vspace{-0.2cm}
 
 If the \betabeta-decay of a given nucleus 
will be observed, it would be possible to 
determine the value of \meff from the
measurement of the associated life-time of the decay.
This would require the knowledge of the nuclear
matrix element of the process. At present there
exist large uncertainties in the calculation of
the \betabeta-decay nuclear matrix elements
(see, e.g., \cite{bb0nu1,bb0nu2}).
This is reflected, in particular, in the
factor of $\sim (2 -3)$ uncertainty
in the upper limit on \meff , which is
extracted from the experimental
lower limits on the \betabeta-decay 
half life-time of $^{76}$Ge. The observation of
a \betabeta-decay of one nucleus is likely to lead
to the searches and eventually to observation
of the decay of other nuclei.
One can expect that such a progress, in particular,
will help to solve completely the problem 
of the sufficiently precise calculation
of the nuclear matrix elements 
for the \betabeta-decay.
Taking the optimistic point of view that
the indicated problem will be resolved 
in one way or another, we will not discuss in what 
follows the possible effects of the currently existing
uncertainties in the evaluation
of the \betabeta-decay nuclear matrix elements
on the results of our analysis.

  In this Section we consider the 
information that future 
\betabeta-decay and/or $^3$H $\beta-$decay 
experiments can provide on the lightest neutrino mass 
$m_1$ and on the CP-violation generated by the two
Majorana CP-violating phases 
$\alpha_{21}$ and  $\alpha_{31}$.
The results are summarized in Fig. 1
(normal neutrino mass hierarchy) and 
in Fig. 2 (inverted hierarchy).

  We shall discuss first the case 
of $\deltasol \equiv \deltadue$
(eqs. (\ref{dmsolnh}) - (\ref{Ue123})). 

\vspace{-0.3cm}
\subsection{Normal Mass Hierarchy: $\deltasol \equiv \deltadue$}

  If $\deltasol = \Delta m^2_{21}$,
for any given solution 
of the solar neutrino problem 
LMA MSW, LOW-QVO, SMA MSW, as can be shown,
\meff can lie anywhere
between 0 and the present upper limits, 
given by eqs. (\ref{76Ge00}) 
and (\ref{IGEX00}).
This conclusion 
does not change even 
under the most favorable
conditions for the determination of \meff,
namely, even when \deltaatm, \deltasol,
$\theta_{\odot}$ and $\theta$ are known
with negligible uncertainty, as Fig. 1 indicates.
The further conclusions that are illustrated
in Fig. 1 are now summarized.
We consider  the case of the 
LMA MSW solution of the solar neutrino problem.

{\bf Case A.} An experimental upper 
limit on \meff, $\meff < \meff_{exp}$,
will determine a maximal value of $m_1$,
 $m_1 < (m_1)_{max}$. The latter 
is fixed by the equality: 
\begin{equation}
(m_1)_{max}: ~~\left| (m_1\cos^2 \theta_\odot - \sqrt{m_1^2 + \deltasol}
\sin^2 \theta_\odot) (1 - |U_{\mathrm{e}3}|^2) \pm
\sqrt{m_1^2 + \deltaatm} |U_{\mathrm{e}3}|^2 
\right| = \meff_{exp}~.
\label{maxm1nhLMA}
\end{equation}
%
\noindent Given $m_1 \neq 0$ and \deltasol,
the sign of the 
last term in the left-hand side 
of the inequality depends 
on the value of 
$\cos 2\theta_\odot$:
the positive sign corresponds to
$\cos 2\theta_\odot < \deltasol \sin^2\theta_\odot / m_1^2 $
(i.e., to $\cos 2\theta_\odot \cong 0$),
while the negative sign is valid
for $\cos 2\theta_\odot > \deltasol \sin^2\theta_\odot / m_1^2 $.

   For the quasi-degenerate neutrino mass spectrum
one has $m_1 \gg \deltasol,\deltaatm$,  
$m_1 \cong m_2 \cong m_3 \cong m_{\nu_e}$, and 
up to corrections 
$\sim \deltasol \sin^2\theta_\odot /(2m_1^2)$
and $\sim \deltaatm|U_{\mathrm{e}3}|^2/(2m_1^2)$ one finds:
\begin{equation}
 (m_1)_{max} \cong \frac{\meff_{exp}}
{\left|\cos 2\theta_\odot (1 - |U_{\mathrm{e}3}|^2) 
- |U_{\mathrm{e}3}|^2 \right|}~.
\label{maxm1nhLMAQD}
\end{equation}
%
\noindent 
If $ |\cos 2\theta_\odot (1 - |U_{\mathrm{e}3}|^2) 
- |U_{\mathrm{e}3}|^2|$ is sufficiently small,
the upper limit on $m_{\nu_e}$ obtained in 
$^3$H $\beta-$decay experiments
could yield  a more 
stringent upper bound  on $m_1$ 
than the bound following from  
the limit on \meff.

{\bf Case B. } 
A measurement  of $\meff = (\meff)_{exp}
\gtap 0.02$ eV
would imply that $m_1 \gtap 0.02$ eV and 
thus a neutrino mass spectrum with partial hierarchy
or of quasi-degenerate type \cite{BPP1}. 
The lightest neutrino mass
will be constrained to lie in the interval,
$(m_1)_{min} \leq m_1 \leq (m_1)_{max}$, 
where $(m_1)_{max}$ and $(m_1)_{min}$
are determined respectively by 
eq. (\ref{maxm1nhLMA})
and by the equation:
\begin{equation}
(m_1)_{min}: ~~
(m_1\cos^2 \theta_\odot + \sqrt{m_1^2 + \deltasol}
\sin^2 \theta_\odot) (1 - |U_{\mathrm{e}3}|^2) +
\sqrt{m_1^2 + \deltaatm} |U_{\mathrm{e}3}|^2 
= (\meff)_{exp}~.
\label{minm1nhLMA}
\end{equation}
%
\noindent 
The limiting values of $m_1$ 
correspond to the case of CP-conservation.
For $\deltasol \ll m_1^2$, \linebreak  (i.e., for 
$\deltasol \ltap 10^{-4}~{\rm eV^2}$),
$(m_1)_{min}$ to a good approximation
is independent of $\theta_{\odot}$,
and for $\sqrt{\deltaatm} |U_{\mathrm{e}3}|^2 \ll m_1$,
which takes place in the case we consider 
as $|U_{\mathrm{e}3}|^2 \ltap 0.05$,
we have $(m_1)_{min} \cong (\meff)_{exp}$.
For $|U_{\mathrm{e}3}|^2 \ll \cos2\theta_{\odot}$,
which is realized in the illustrative cases in 
Fig. 1 for $|U_{\mathrm{e}3}|^2 \ltap 0.01$, 
practically all of the region between
$(m_1)_{min}$ and $(m_1)_{max}$,
$(m_1)_{min} < m_1 < (m_1)_{max}$,
corresponds to violation of the CP-symmetry.
If $|U_{\mathrm{e}3}|^2$ is non-negligible
with respect to $\cos2\theta_{\odot}$,
e.g., if $|U_{\mathrm{e}3}|^2 \cong (0.02 - 0.05)$
for the values of $\cos2\theta_{\odot}$
used to derive the 
right panels in Fig. 1~,
one can have $(m_1)_{min} < m_1 < (m_1)_{max}$
if CP-symmetry is violated, as well as  
in  two specific cases of CP-conservation.
One of these two CP-conserving values of $m_1$, 
corresponding to $\eta_{21} = - \eta_{31} = -1$,
can differ considerably from the two limiting values
(see Fig. 1). 

 A measured value of \meff satisfying 
$(\meff)_{exp} < (\meff)_{max}$, where 
$(\meff)_{max}$ $ \cong m_1 \cong m_{\nu_e}$
in the case of a quasi-degenerate 
neutrino mass spectrum, and 
$(\meff)_{max} \cong 
(\sqrt{\deltasol}
\sin^2 \theta_\odot) (1 - |U_{\mathrm{e}3}|^2) +
\sqrt{\deltaatm} |U_{\mathrm{e}3}|^2$ if 
the spectrum is hierarchical 
(i.e., if $m_1 \ll m_2 \ll m_3$), 
would imply that 
at least one of the two CP-violating phases
is different from zero :
$\alpha_{21}\neq 0$ or/and $\alpha_{31} \neq 0$;
in the case of a hierarchical spectrum
that would also imply $\alpha_{21}\neq\alpha_{31}$.
In general, the knowledge of the value
of \meff alone will not allow to 
distinguish the case
of CP-conservation 
from that of CP-violation.

{\bf Case C.} 
It might be possible to 
determine whether
CP-violation due to the Majorana 
phases takes place in the lepton sector if 
both \meff and $m_{\nu_e}$ are measured.
Since prospective measurements 
are limited to $(m_{\nu_e})_{exp} \gtap 0.35$ eV,
the relevant neutrino mass spectrum is of 
quasi-degenerate type (see, e.g., \cite{BPP1}). 
In this case one has
$m_1 > 0.35$ eV,
$m_1 \cong m_2 \cong m_3 \cong m_{\nu_e}$ and
\begin{equation}
\meff \simeq m_{\nu_e} 
\left| \cos^2 \theta_\odot (1- \uetre) 
+ \sin^2    \theta_\odot (1- \uetre)e^{i\alpha_{21}}  
+ |U_{\mathrm{e} 3}|^2 e^{i \alpha_{31}} \right|.
\label{meffmdeg}
\end{equation}
%
\noindent If we can neglect $|U_{\mathrm{e}3}|^2$ 
in eq. (\ref{meffmdeg}) (i.e., 
if $\cos 2\theta_\odot \gg |U_{\mathrm{e}3}|^2$),
a value of $m_{\nu_e} \cong m_1$, satisfying
$(m_1)_{min} < m_{\nu_e} < (m_1)_{max}$,
where $(m_1)_{min}$ and $(m_1)_{max}$ are determined by
eqs. (\ref{minm1nhLMA}) and (\ref{maxm1nhLMA}),
would imply that the CP-symmetry 
does not hold in the lepton sector. 
In this case one would obtain 
correlated constraints
on the CP-violating phases 
$\alpha_{21}$ and $\alpha_{31}$ 
\cite{BPP1,Rodej00}.
This appears to be the only possibility 
for demonstrating   
CP-violation  due to Majorana  
CP-violating phases in the case of
$\deltasol \equiv \deltadue$ under discussion.
In order to reach a definite conclusion
concerning CP-violation  due to the Majorana  
CP-violating phases, considerable accuracy
in the measured values of
\meff and $m_{\nu_e}$ is required.
For example, if the oscillation experiments
give the result $\cos2\theta_{\odot} \leq 0.3$
and $\meff = 0.3$ eV, a value of 
$m_{\nu_e}$  between 0.3 eV and 1.0 eV
would demonstrate CP-violation.
However, this requires better
than 30\% accuracy on both measurements.
The accuracy requirements become
less stringent if the upper limit on 
$\cos2\theta_{\odot}$ is smaller. 
 
  If $\cos2\theta_{\odot} > |U_{\mathrm{e}3}|^2$ but 
$|U_{\mathrm{e}3}|^2$ cannot be neglected
in (\ref{meffmdeg}),  
there exist two 
CP-conserving values of  $m_{\nu_e}$
in the interval $(m_1)_{min} < m_{\nu_e} < (m_1)_{max}$.
The one that can significantly differ 
from the extreme values of the interval
corresponds to a specific case of 
CP-conservation - to 
$\eta_{21} = - \eta_{31} = -1$ (Fig. 1).

{\bf Case D. } 
A measured value of $m_{\nu_e}$,
$(m_{\nu_e})_{exp} \gtap 0.35$ eV, satisfying 
$(m_{\nu_e})_{exp} > (m_1)_{max}$,
where $(m_1)_{max}$ is determined from the upper limit
on \meff, eq. (\ref{maxm1nhLMA}), in the case
the \betabeta-decay is not 
observed, might imply 
that the massive neutrinos are Dirac particles.
If \betabeta-decay
 has been observed and  \meff 
measured, the inequality
$(m_{\nu_e})_{exp} > (m_1)_{max}$, with
$(m_1)_{max}$ determined from 
the upper limit or the value of \meff,
eq. (\ref{maxm1nhLMA}), 
would lead to the conclusion that
there exist contribution(s) to
the \betabeta-decay rate other than 
due to the light Majorana neutrino exchange
(see, e.g., \cite{bb0nunmi} 
and the references quoted therein) that 
partially cancels the
contribution from the 
Majorana neutrino exchange.

 A measured value of \meff, 
$( \meff)_{\mathrm{exp}} \gtap 0.01 \  \mathrm{eV}$,
and a  measured value of 
$m_{\nu_e}$ or an  upper bound on $m_{\nu_e}$
such that $m_{\nu_e} < (m_1)_{min}$,
where $(m_1)_{min}$ is determined by 
eq. (\ref{minm1nhLMA}), 
would imply that there are contributions to the
\betabeta-decay rate in addition to the ones
due to the light Majorana neutrino exchange
(see, e.g., \cite{bb0nunmi}), 
which enhance the \betabeta-decay rate
and signal the existence of new $\Delta L =2$
processes beyond those induced 
by the light Majorana neutrino exchange
in the case of left-handed charged current  weak 
interaction.

{\bf Case E.} 
An actual measurement of 
$\meff \ltap 10^{-2}$ eV is unlikely, 
but it is illustrated in Fig. 1  to show
the interpretation of such a result. 
There always remains an upper limit on $m_1$.
As \meff decreases, 
there appears a finite lower limit on $m_1$
as well. Both the upper and the lower
limits  on $m_1$ 
approach asymptotic values which 
depend on the values of \deltasol,
\deltaatm, $\cos2\theta_{\odot}$ 
and $|U_{e3}|^2$, but are 
independent of \meff (Fig. 1).
For $\cos2\theta_{\odot} \gtap 2|U_{e3}|^2$, 
the maximum and minimum asymptotic values of 
$m_1$ are determined by the expressions:
$$
m_1^{(\pm)} = 
\left (- \eta_{31}\sqrt{\deltaatm} 
|U_{e3}|^2\cos^2\theta_{\odot} \right. ~~~~~~~~~~~~~~~~~~~~~~~~~~~~~~~~~$$
\begin{equation}
~~~~~~~~~~~\left. \pm ~\sqrt{\deltaatm |U_{e3}|^4\cos^4\theta_{\odot} -
(\deltaatm |U_{e3}|^4 - \deltasol \sin^4\theta_{\odot})
\cos2\theta_{\odot}} \right )\cos^{-1}2\theta_{\odot}.
\label{amaxminm1}
\end{equation}
%
\noindent For the maximum asymptotic value we have
$(m_1)_{max} = m_1^{(+)}$ with $\eta_{31} = - 1$.
If further $\deltaatm |U_{e3}|^4$
$\times \cos^4\theta_{\odot} \gg
|(\deltaatm |U_{e3}|^4 - \deltasol \sin^4\theta_{\odot})
\cos2\theta_{\odot}|$ (which 
requires $|U_{e3}|^2 \cong (0.02 - 0.05)$),
the expression for the asymptotic value of interest 
is given approximately by
$(m_1)_{max} \cong 2\sqrt{\deltaatm} |U_{e3}|^2$
$\times \cos^2\theta_{\odot}/\cos2\theta_{\odot}$
and is typically in the range
$(m_1)_{max} \cong (0.7 - 3.0)\times 10^{-2}$ eV
(Fig. 1, right panels).
If, however, 
$\deltasol \sin^4\theta_{\odot} \gg
{\rm max}(\deltaatm |U_{e3}|^4,
\deltaatm |U_{e3}|^4\cos^4\theta_{\odot}/\cos2\theta_{\odot})$,
one typically finds: 
$(m_1)_{max} \cong (0.3 - 1.0)\times 10^{-2}$ eV
(Fig. 1, left panels).

  For the minimum asymptotic value of $m_1$ 
we have $(m_1)_{min} = m_1^{(+)}$
with $\eta_{31} = 1$
if $\deltasol\sin^4\theta_{\odot} > 
\deltaatm |U_{e3}|^4$, and
$(m_1)_{min} = m_1^{(-)}$ with 
$\eta_{31} = - 1$
if $\deltasol \sin^4\theta_{\odot} <
\deltaatm |U_{e3}|^4$.

  Over certain interval of values of \meff,
which depends on $|U_{e3}|^2$,
on the values of the difference of the
Majorana CP-violating phases,
$(\alpha_{31} - \alpha_{21})$,
and on $\cos2\theta_{\odot}$,
the lower limit on $m_1$ goes to zero,
as is shown in Fig. 1. This interval,
$\meff_{-}~ \leq \meff \leq \meff_{+}$,
is given by 
$\meff_{\pm} = |\sqrt{\deltasol}
\sin^2 \theta_\odot (1 - |U_{\mathrm{e}3}|^2) \pm
\sqrt{\deltaatm} |U_{\mathrm{e}3}|^2|$,
and has a width 
of $2\sqrt{\deltaatm} |U_{\mathrm{e}3}|^2|$. 
 For a given \meff from the indicated 
interval we have 
$0 \leq m_1 \leq (m_1)_{max}$,
with $(m_1)_{max}$ determined by 
eq. (\ref{maxm1nhLMA}). Further,
the limiting value of 
$m_1 = (m_1)_{max}$,
as well as at least one
and up to three internal values 
of $m_1$ from the interval
$0 < m_1 < (m_1)_{max}$ in the simplified
case we are analyzing
are CP-conserving (Fig. 1).
The remaining values of 
$m_1$ from the interval
$0 \leq m_1 < (m_1)_{max}$
are CP-violating.

  It should be noted also that 
one can have $\meff = 0$ for $m_1 = 0$
in the case of CP-invariance if 
$\eta_{21} = - \eta_{31}$
and the relation
$\sqrt{\deltasol}
\sin^2 \theta_\odot (1 - |U_{\mathrm{e}3}|^2) =
\sqrt{\deltaatm} |U_{\mathrm{e}3}|^2$ holds.
Finally, there would seem to be no practical possibility
to determine the Majorana CP-violating phases.

   The analysis of the {\it Cases A - E} 
for the LOW-QVO solution of the     
solar neutrino problem leads to the same
qualitative conclusions as those obtained above
for the LMA MSW solution. 
The conclusions differ,
however, in the case of the SMA MSW solution 
and we will discuss them next briefly.
An experimental upper limit on 
\meff ({\it Case A}) in the range
$\meff_{exp} \geq 10^{-2}$ eV, would imply 
in the case of the SMA MSW solution, 
$m_1 < \meff_{exp}(1 - 2|U_{e3}|^2)^{-1}$.
For values of $\meff \gtap 10^{-2}$ eV,
the maximum and minimum values of 
$m_1$ are extremely close:
$(m_1)_{min} \cong \meff_{exp}$.
As a result, 
a measurement of \meff ({\it Case B}) 
practically determines $m_1$,
$m_1 \cong \meff$. However,  
there is no possibility 
to determine or constrain
the Majorana CP-violating
phases. Thus, no information about CP-violation
generated by the Majorana phases 
can be obtained by the measurement of
\meff  (or of \meff and $m_{\nu_e}$) 
\cite{BPP1}. If both 
$\meff \gtap 0.02$ eV and 
$m_{\nu_e}\gtap 0.35$ eV 
would be measured ({\it Case C}),
the relation $m_1 \cong (\meff)_{exp} 
\cong (m_{\nu_e})_{exp}$
should hold. The conclusions in the 
{\it Cases D} and {\it E}
are qualitatively the same as for the LMA MSW 
solution.

\vspace{-0.2cm}
\subsection{Inverted Mass Hierarchy: $\deltasol \equiv \deltatre$}

  Consider next the possibility of a 
neutrino mass spectrum with inverted 
hierarchy, which is illustrated in Fig. 2. 
A comparison of Fig. 1 and Fig. 2
reveals two major differences 
in the predictions for \meff:
if $\deltasol \equiv \deltatre$, 
i) even in the case of $m_1 \ll m_2 \cong m_3$
(i.e., even if $m_1 \ll 0.02$ eV), \meff can 
exceed $\sim 10^{-2}$ eV
and can reach the value of $\sim 0.08$ eV 
\cite{BGGKP99,BPP1}, and 
 ii) a more precise determination of 
\deltaatm, \deltasol, 
$\theta_{\odot}$ 
and $\sin^2\theta = |U_{e1}|^2$, 
can lead to a lower limit on the possible values of 
\meff  \cite{BPP1}.
For the LMA and the LOW-QVO
solutions, ${\rm min}(\meff)$
will depend, in particular,
on whether CP-invariance 
holds or not in the lepton sector, 
and if it holds -
on the relative CP-parities 
of the massive Majorana neutrinos.
All these possibilities are parametrized
by the values of the two  
CP-violating phases,
$\alpha_{21}$ and $\alpha_{31}$,
entering into the expression for \meff.
The existence of a significant lower
limit on the possible values
of \meff depends crucially in the cases of the
LMA and LOW-QVO solutions on the
minimal value of $|\cos2\theta_{\odot}|$, 
allowed by the data: up to corrections
$\sim 5\times 10^{-3}$ eV we have for these
two solutions (see, e.g., \cite{BGKP96,BGGKP99,BPP1}):
%
\begin{equation}
{\rm LMA,~LOW-QVO}:~{\rm min}(\meff)_{LMA} \cong \left| \sqrt{\deltaatm}
|\cos2\theta_{\odot}|(1 - |U_{\mathrm{e} 1}|^2) \pm 
0(\sim 5\times 10^{-3}~{\rm eV}) \right|.~~
\label{minmeffLA}
\end{equation}
%
\noindent The min(\meff) in eq. (\ref{minmeffLA}) 
is reached in the case of CP-invariance and
$\eta_{21} = - \eta_{31} = \pm 1$. 
If $\cos2\theta_{\odot} = 0$ 
is allowed,
values of \meff smaller than
$\sim 5\times 10^{-3}$ eV and even 
$\meff = 0$ would 
be possible. 
If, however, it will be 
experimentally established that, e.g.,
$|\cos2\theta_{\odot}|\gtap 0.20$,
we will have 
${\rm min}(\meff) \cong 0.01$ eV
if \deltaatm and $|U_{\mathrm{e} 1}|^2$
lie within their
$90\% $ C.L. allowed regions found
in \cite{Gonza3nu}~
\footnote{If, for instance,
$|\cos2\theta_{\odot}|\gtap 0.30;~0.50$,
then under the same conditions
one will have 
${\rm min}(\meff) \cong 0.015;~0.025$ eV.}
(i.e., $|U_{\mathrm{e} 1}|^2 < 0.055$, 
$\deltaatm = (1.4 - 6.1)\times 10^{-3}~{\rm eV^2}$).
According to the 
latest analysis of the solar neutrino data
(including the SNO results) performed in
\cite{FogliSNO}, for the LMA MSW solution
one has $\cos2\theta_{\odot}\gtap 0.30~(0.50)$
at 99\% (95\%) C.L. 
 
  For the SMA MSW solution one has
in the case of $\deltasol = \Delta m^2_{32}$
under discussion: 
%
\begin{equation}
{\rm SMA~ MSW}:~~~~{\rm min}(\meff)_{SMA} \cong \meff 
\cong \left| \sqrt{\deltaatm}
(1 - |U_{\mathrm{e} 1}|^2) \pm 
0(\sim 5\times 10^{-3}~{\rm eV}) \right|,~~~~
\end{equation}
%
\noindent where $|U_{\mathrm{e} 1}|^2$ is 
limited by the CHOOZ data. 
Using the current 99\%~(90\%) C.L.
allowed values of \deltaatm
and $|U_{\mathrm{e} 1}|^2$, 
derived  in \cite{Gonza3nu}, one finds
${\rm min}(\meff) \cong 0.030~(0.050)$ eV.

 We shall discuss next briefly 
the implications of the results
of future \betabeta-decay and 
\hbeta experiments.
We follow the same line of analysis 
we have used for neutrino mass spectrum 
with normal hierarchy. Consider the case 
of the LMA MSW solution 
of the solar neutrino problem. 

{\bf Case A.} An experimental upper 
limit on \meff, $\meff < \meff_{exp}$,
which is larger than the 
minimal value of \meff, 
$\meff_{min}^{ph}$, 
predicted by taking into account 
all uncertainties in the values of 
the relevant input parameters 
(\deltaatm, \deltasol, $\theta_{\odot}$, etc.),
$\meff_{exp} \geq \meff_{min}^{ph}$,
will imply an upper limit on $m_1$, $m_1 < (m_1)_{max}$.
The latter is determined by the equality: 
%
$$ (m_1)_{max}:~~~~\left | \left ( \sqrt{m_1^2 + \deltaatm - \deltasol}
\cos^2 \theta_\odot \right. \right.~~~~~~~~~~~~~~~~~~~~~~~~~~~~~~~~~~~~~~~~~~~~~~~~$$ 
\begin{equation}
~~~~~~~~~~~~~~~~~~~~~~~ \left. \left. - ~~\sqrt{m_1^2 + \deltaatm}
\sin^2 \theta_\odot \right ) (1 - |U_{\mathrm{e}1}|^2)
\pm m_1 |U_{\mathrm{e}1}|^2 \right |  = \meff_{exp}~.
\label{maxm1ih}
\end{equation}
%
\noindent The term $m_1 |U_{\mathrm{e}1}|^2$ 
enters with a plus (minus) sign if the 
difference between the two terms
in the big round brackets in the
left-hand side of the equation is negative
(positive).
For the quasi-degenerate neutrino mass spectrum
($m_1 \gg \deltasol,\deltaatm$,  
$m_1 \cong m_2 \cong m_3 \cong m_{\nu_e}$), 
$(m_1)_{max}$ is given by eq. (\ref{maxm1nhLMAQD})
in which $|U_{\mathrm{e}3}|^2$ is replaced by 
$|U_{\mathrm{e}1}|^2$.
Correspondingly, the conclusion that if 
$|\cos 2\theta_\odot (1 - |U_{\mathrm{e}1}|^2) 
- |U_{\mathrm{e}1}|^2|$ is 
sufficiently small,
the upper limit on $m_1 \cong m_{\nu_e}$, 
obtained in $^3$H $\beta-$decay, can 
be more stringent than 
the upper bound  on $m_1$, 
implied by the limit on \meff,
remains valid.

   An experimental upper 
limit on \meff,
which is smaller than the 
minimal possible value of \meff, 
$\meff_{exp} < \meff_{min}^{ph}$,
would imply that either 
i) the neutrino mass spectrum is not
of the inverted hierarchy type, or ii) that 
there exist contributions to
the \betabeta-decay rate other than 
due to the light Majorana neutrino exchange
(see, e.g., \cite{bb0nunmi}) that 
partially cancel the
contribution from the 
Majorana neutrino exchange. 
The indicated result might also 
suggest that the massive neutrinos 
are Dirac particles.

{\bf Case B.} 
A measurement of $\meff = (\meff)_{exp}
\gtap \sqrt{\deltaatm}(1 -  |U_{\mathrm{e}1}|^2)
\cong (0.04  - 0.08)$ eV, where we have used the 
90\% C.L. allowed regions of \deltaatm and  
$|U_{\mathrm{e}1}|^2$ from \cite{Gonza3nu},
would imply the existence of a finite 
interval of possible values of $m_1$,
$(m_1)_{min} \leq m_1 \leq (m_1)_{max}$, 
with $(m_1)_{max}$ and $(m_1)_{min}$
given respectively by eq. (\ref{maxm1ih}) and by 
%
$$ (m_1)_{min}:~~~~~ m_1 |U_{\mathrm{e}1}|^2 + 
\left ( \sqrt{m_1^2 + \deltaatm - \deltasol}
\cos^2 \theta_\odot \right.~~~~~~~~~~~~~~~~~~~~~~~~~~~~~~~~~~~~~~~~~~~~ $$ 
\begin{equation}
~~~~~~~~~~~~~~~~~~~~~~~\left. +~\sqrt{m_1^2 + \deltaatm}
\sin^2 \theta_\odot \right ) (1 - |U_{\mathrm{e}1}|^2) 
= \meff_{exp}~.
\label{minm1ih}
\end{equation}
%
\noindent In this case $m_1 \gtap 0.04$ eV and 
the neutrino mass spectrum is with partial inverted 
hierarchy or of quasi-degenerate type \cite{BPP1}. 
The limiting values of $m_1$ 
correspond to CP-conservation.
For $\deltasol \ll m_1^2$, i.e.,
for $\deltasol \ltap 10^{-4}~{\rm eV^2}$,
$(m_1)_{min}$ is to a good approximation
independent of $\theta_{\odot}$ and
we have: $\sqrt{((m_1)_{min})^2 + \deltaatm}~
(1 - |U_{\mathrm{e}1}|^2)\cong (\meff)_{exp}$.

   For negligible $|U_{\mathrm{e}1}|^2$
(i.e., $|U_{\mathrm{e}1}|^2 \ltap 0.01$ 
for the values of $\cos2\theta_{\odot}$ in Fig. 2),
essentially all of the interval between
$(m_1)_{min}$ and $(m_1)_{max}$,
$(m_1)_{min} < m_1 < (m_1)_{max}$,
corresponds to violation of the CP-symmetry.
If the terms $\sim |U_{\mathrm{e}1}|^2$ cannot be neglected 
in eqs. (\ref{maxm1ih}) and (\ref{minm1ih}) 
(i.e., $|U_{\mathrm{e}1}|^2 \cong (0.02 - 0.05)$
for the values of $\cos2\theta_{\odot}$ in Fig. 2),
there exists for a fixed $\meff_{exp}$
two CP-conserving values of $m_1$ in the  
indicated interval, one of which differs
noticeably from the limiting values 
$(m_1)_{min}$ and  $(m_1)_{max}$
and corresponds to 
$\eta_{21} = - \eta_{31} = 1$ (Fig. 2).

 In general, measuring the value
of \meff alone will not allow to 
distinguish the case
of CP-conservation from that of CP-violation.
In principle, a measurement of 
$m_{\nu_e}$, or even an upper limit 
on $m_{\nu_e}$, smaller than $(m_1)_{max}$,
could be a signal of CP-violation.
However, unless $\cos2\theta_{\odot}$
is very small, the required values of 
$m_{\nu_e}$ are less than
prospective measurements. For example,
as is seen in Fig. 2, upper left panel,
for $\cos2\theta_{\odot} = 0.1$ and
$\meff = 0.03$ eV, one needs to find
$m_{\nu_e} < 0.35$ eV to demonstrate
CP-violation.

  If the measured value of \meff lies 
in the interval
$\meff_{-}~ \leq \meff \leq \meff_{+}$, where
%
\begin{equation}
\meff_{\pm} = 
\left |\sqrt{\deltaatm - \deltasol}
\cos^2 \theta_\odot \pm \sqrt{\deltaatm}
\sin^2 \theta_\odot \right | (1 - |U_{\mathrm{e}1}|^2), 
\label{0m1ih}
\end{equation}
%
\noindent we would have $(m_1)_{min} = 0$.
The values of $m_1$ satisfying
$0 \leq m_1 < (m_1)_{max}$, where 
$(m_1)_{max}$ is determined by eq. (\ref{maxm1ih}),
correspond to violation of the CP-symmetry (Fig. 2).

  {\bf Cases C.} As Fig. 2 indicates,
the discussions 
and conclusions are identical to 
the discussions and conclusions in the same
cases for the neutrino mass spectrum 
with normal hierarchy, except that
instead of eq. (20) we have 
\begin{equation}
\meff \simeq m_{\nu_e} 
\left||U_{\mathrm{e} 1}|^2 + \cos^2 \theta_\odot (1- \ueuno) 
e^{i\alpha_{21}}
+ \sin^2 \theta_\odot (1- \ueuno)e^{i \alpha_{31}} \right|,
\label{meffmdegih}
\end{equation}
%
\noindent $(m_1)_{max}$ and $(m_1)_{min}$
are determined by eqs. (\ref{maxm1ih})
and (\ref{minm1ih}), and  
$|U_{\mathrm{e}3}|^2$ must be substituted
by $|U_{\mathrm{e}1}|^2$ in the 
relevant parts of the analysis.

  {\bf Case D.}
 If $m_{\nu_e}$ is measured 
and  $(m_{\nu_e})_{exp} \gtap 0.35 \ \mathrm{eV}$ 
but the \betabeta-decay is not observed 
or is observed and
$(m_{\nu_e})_{exp} > (m_1)_{max}$,
where $(m_1)_{max}$ is determined by eq. (\ref{maxm1nhLMA}),
the same considerations and conclusions as
 in the Case D for the normal hierarchy
mass spectrum apply.

A measured value of \meff, 
$( \meff)_{\mathrm{exp}} \gtap 0.1 \ \mathrm{eV}$,
in the case when the measured value of 
$m_{\nu_e}$ or the upper bound on $m_{\nu_e}$
are such that $m_{\nu_e} < (m_1)_{min}$,
where  $(m_1)_{min}$ is determined by 
eq. (\ref{minm1ih}), 
would lead to the 
same conclusions as  in the Case D
for the normal hierarchy
mass spectrum. 

  {\bf Case E.} 
It is possible to have a 
measured value of 
$\meff \ltap 10^{-2}$ eV 
in the case of 
the LMA MSW solution and 
neutrino mass spectrum
with inverted hierarchy 
under discussion only if 
$\cos2\theta_{\odot}$ is 
rather small, $\cos2\theta_{\odot}\ltap 0.2$.
A measured value of 
$\meff < \meff_{min}^{ph}$ would imply
that either
the neutrino mass spectrum is not
of the inverted hierarchy type, or  that 
there exist contributions to
the \betabeta-decay rate other than 
due to the light Majorana neutrino exchange
that partially cancel the
contribution from the 
Majorana neutrino exchange.

   The above conclusions hold with minor 
modifications (essentially of the numerical 
values involved) for the 
LOW-QVO solution as well. In the case of the 
SMA MSW solution we have, as is well-known, 
$\sin^2\theta_{\odot} << 1$ and 
$\deltasol \ltap 10^{-5}~{\rm eV^2}$ (see, e.g.,
\cite{ConchaSNO}). Consequently, the analog
of eq. (\ref{maxm1nhLMAQD}) in {\it Case A} reads
$(m_1)_{max} \cong \meff_{exp}
(1 - 2|U_{\mathrm{e}1}|^2)^{-1}$.
The conclusions in the  {\it Cases B - D}
are qualitatively the same 
as in the case of
neutrino mass spectrum with normal hierarchy.
In particular, a measured value of 
$\meff > \meff_{+} \cong \sqrt{\deltaatm}
(1 - |U_{\mathrm{e} 1}|^2)$, would essentially
determine $m_1$, $m_1 \cong (\meff)_{exp}$.
No information about CP-violation
generated by the Majorana phases 
can be obtained by the measurement of
\meff, or of \meff and $m_{\nu_e}$.
If both $\meff$ and  $m_{\nu_e}\gtap 0.35$ eV 
are measured, the relation $m_1 \cong (\meff)_{exp} 
\cong (m_{\nu_e})_{exp}$
should hold.
If it is found that 
$\meff = \sqrt{\deltaatm} (1 - |U_{\mathrm{e} 1}|^2)$,
one would have 
$0 \leq m_1 \leq (m_1)_{max}$, where 
$(m_1)_{max}$ is determined by eq. (\ref{maxm1ih})
in which effectively  
$\sin^2\theta_{\odot} = 0$,
$\cos^2\theta_{\odot} = 1$,
and $\deltasol = 0$.
Finally, a measured value of 
$\meff < \meff_{-} \cong \meff_{+} \cong \sqrt{\deltaatm}
(1 - |U_{\mathrm{e} 1}|^2)$
would either indicate that 
there exist new additional contributions 
to the \betabeta-decay rate,
or that the SMA MSW solution
is not the correct solution of the
solar neutrino problem.

\vspace{-0.2cm}
\section{Conclusions}

  Neutrino oscillation experiments can never 
tell the actual neutrino masses (that is, the lowest mass $m_1$), 
whether neutrinos are Majorana, and, if so, 
whether there are Majorana 
CP-violating phases associated with 
the $\Delta L=2$ neutrino mass. 
Neutrinoless double-beta 
decay experiments can, in principle, 
answer the first two questions, 
but cannot by themselves provide information
about CP-violation. 
Here we have analyzed how, 
given optimum information 
from neutrino oscillation 
and \betabeta-decay experiments,  
a measurement of neutrino mass 
from \hbeta-decay could, in principle, 
give evidence for Majorana
CP-violating phases, even though no 
CP-violation would be directly observed. 
The indicated possibility requires quite 
accurate measurements and holds only for a 
limited range of parameters.  
          
\vspace{0.2cm}
\noindent {\bf Note Added.} After the completion of 
the present paper we became aware of the very recent work 
\cite{Poles2}, where some of the topics we discuss 
are also considered but within a somewhat different
approach.

\vspace{0.2cm}
\noindent {\bf Acknowledgements} 
The work of L.W. was carried out in part at the 
Aspen Center for Physics and was supported in part 
by the U.S. Department of Energy under grant 
DE-FG02-91ER407682.
S.T.P. would like to acknowledge with gratefulness
the hospitality and support of the SLAC Theoretical 
Physics Group, where part of the work on 
the present study was done. 
The work of S.T.P. was supported in part by the EEC grant 
ERBFMRXCT960090.
S.P. would like to thank the Theoretical Physics Group 
of the University of Sussex where part of this work was done.

\vspace{-0.2cm}

\vspace{0.3cm}


\vspace{0.2cm}
\begin{figure}[t]
\begin{center}
\epsfig{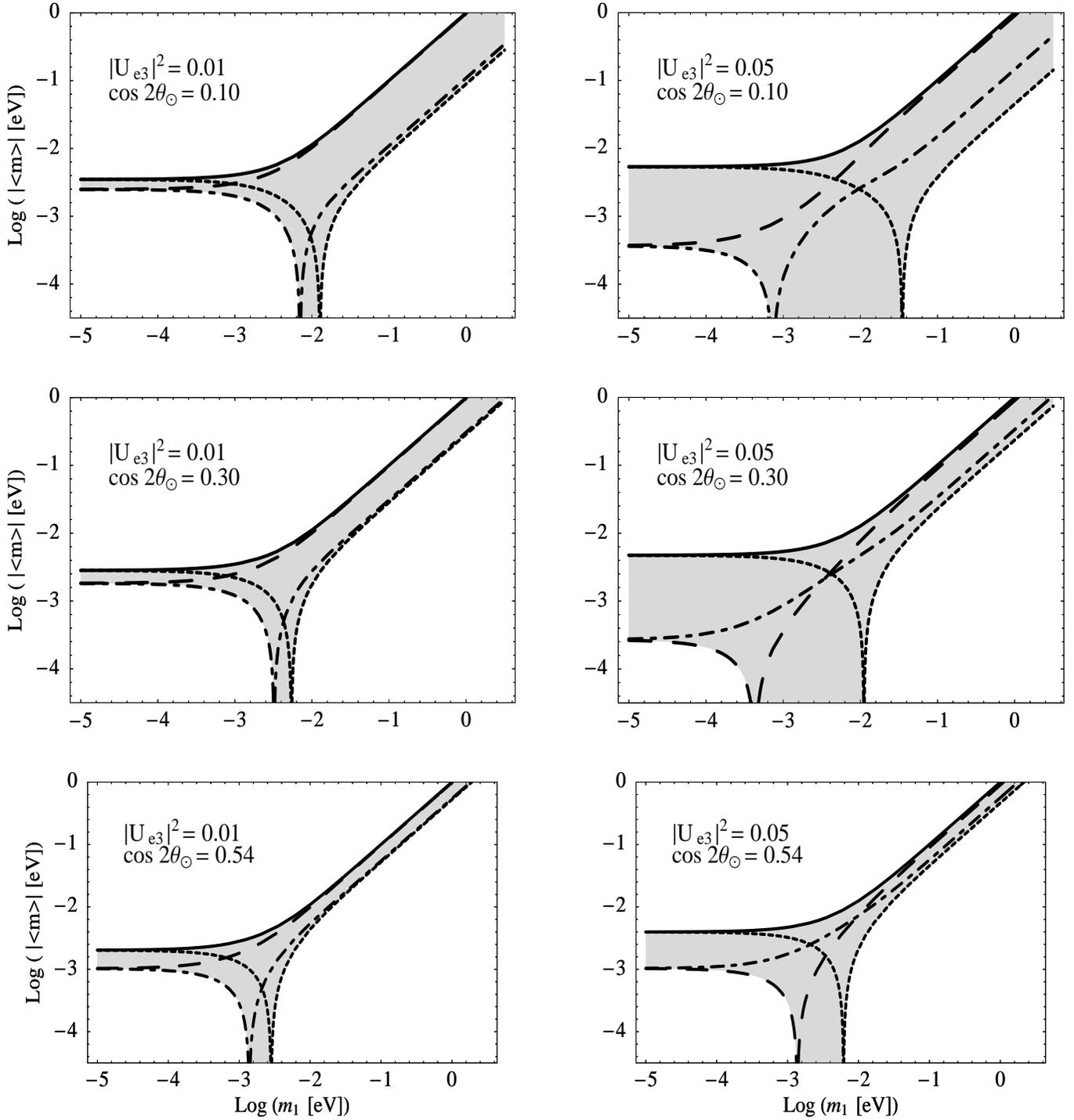}
\end{center}
\caption{The dependence of \meff on $m_1$ 
in the case of $\deltasol = \Delta m_{21}^2$
(normal hierarchy of neutrino masses)
for the LMA MSW solution 
of the solar neutrino problem.
The three vertical left (right) panels correspond 
to $|U_{\mathrm{e} 3}|^2 = 0.01~(0.05)$,
while the two upper, the two middle 
and the two lower
panels are obtained respectively for
$\cos2\theta_{\odot} = 0.10;~0.30;~0.54$.
The figures are obtained for the best fit values 
of \deltaatm and \deltasol, given in 
eqs. (\ref{bfdmatm}) and (\ref{bfdmsol}).
In the case of CP-conservation
the allowed values of \meff are constrained to
lie on i) the solid line if
$\eta_{21} = \eta_{31} = 1$,
ii) on the dashed line if
$\eta_{21} = - \eta_{31} = 1$,
iii) on the dotted lines 
if $\eta_{21} =  \eta_{31} = - 1$, and
iv) on the dash-dotted lines
if $\eta_{21} = - \eta_{31} = -1$.
The region colored in grey  
(not including these lines) 
requires CP-violation
(``just CP-violation'' region).
} 
\label{figure:massdeg00}
\end{figure}
\begin{figure}[t]
\begin{center}
\epsfig{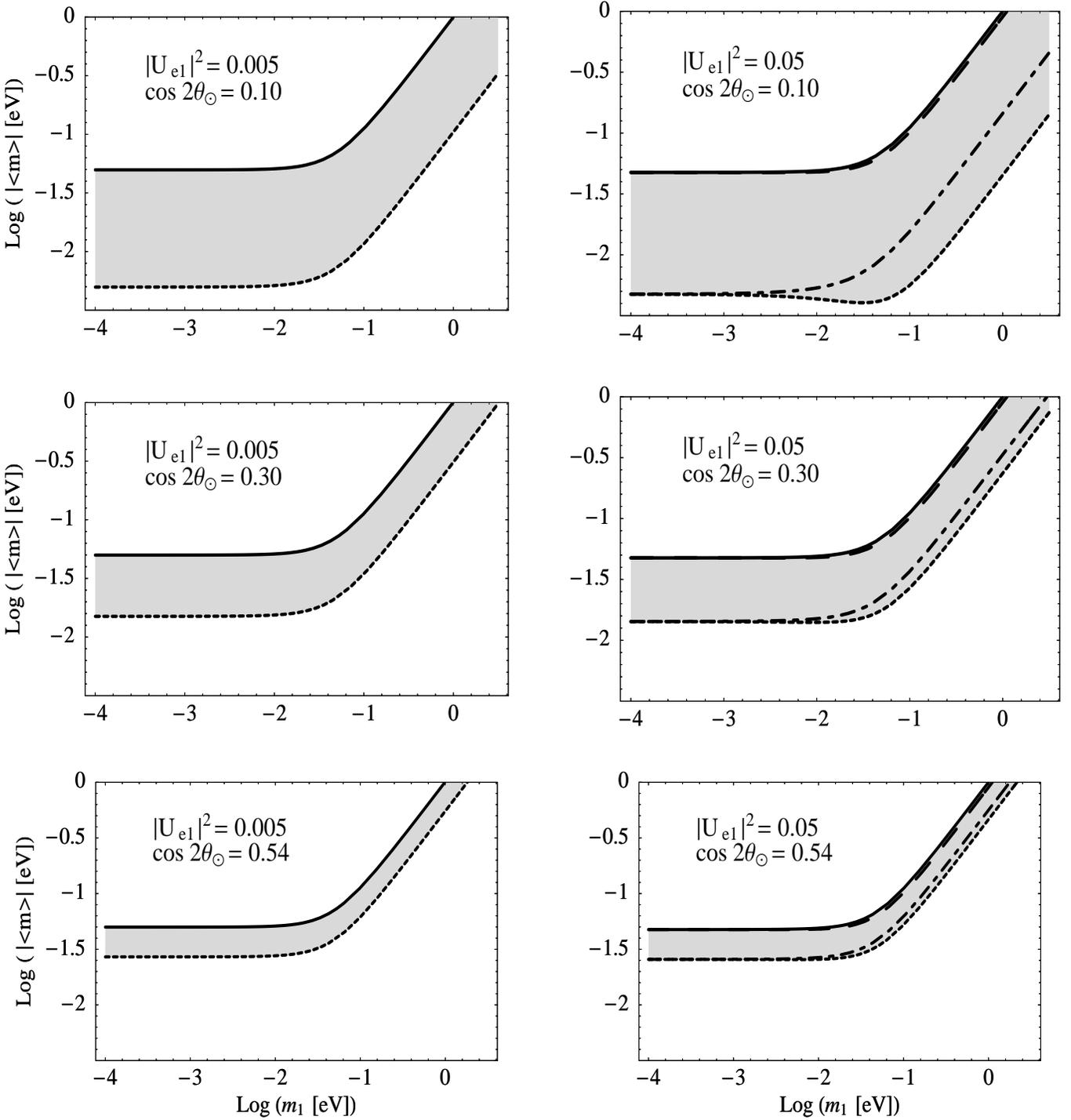}
\end{center}
\caption{The same as Fig. 1 for the inverted hierarchy,
$\deltasol = \Delta m_{32}^2$.
The three vertical left (right) panels correspond 
to $|U_{\mathrm{e} 3}|^2 = 0.005~(0.05)$,
while the two upper, the two middle 
and the two lower
panels are obtained respectively for
$\cos2\theta_{\odot} = 0.10;~0.30;~0.54$.
The figures are obtained for the best fit values 
of \deltaatm and \deltasol, given in 
eqs. (\ref{bfdmatm}) and (\ref{bfdmsol}).
If CP-invariance holds,
the allowed values of \meff are constrained to
lie on i) the solid line if
$\eta_{21} = \eta_{31} = 1$,
ii) on the dashed line if
$\eta_{21} = \eta_{31} = -1$,
iii) on the dotted line
if $\eta_{21} =  - \eta_{31} = - 1$, and
iv) on the dash-dotted lines
if $\eta_{21} = - \eta_{31} = 1$
for $|U_{\mathrm{e} 3}|^2 = 0.05$ 
and on v) the solid line if
$\eta_{21} = \eta_{31} = \pm 1$,
vi) on the dotted line
if $\eta_{21} =  - \eta_{31} = \pm 1$  
for $|U_{\mathrm{e} 3}|^2 = 0.005$.
The region colored in grey 
(not including the indicated lines)
requires CP-violation.
} 
\label{figure:massdeg00}
\end{figure}

\end{document}